\documentstyle[12pt]{article}

\advance \topmargin by -\headheight
\advance \topmargin by -\headsep

\evensidemargin \oddsidemargin
\begin{document}

\topmargin -.6in

%               macros formatting and equations
\def\rf#1{(\ref{eq:#1})}
\def\lab#1{\label{eq:#1}}
\def\br{\begin{eqnarray}}
\def\er{\end{eqnarray}}
\def\be{\begin{equation}}
\def\ee{\end{equation}}
\def\nn{\nonumber}
\def\lb{\lbrack}
\def\rb{\rbrack}
\def\({\left(}
\def\){\right)}
\def\v{\vert}
\def\bv{\bigm\vert}
\def\lskip{\vskip\baselineskip\vskip-\parskip\noindent}
\relax
\newcommand{\nit}{\noindent}
\newcommand{\ct}[1]{\cite{#1}}
\newcommand{\bi}[1]{\bibitem{#1}}
%%                                math symbols
%
\def\a{\alpha}
\def\b{\beta}
\def\c{\chi   }
\def\ca{{\cal A}}
\def\cm{{\cal M}}
\def\cn{{\cal N}}
\def\cf{{\cal F}}
\def\d{\delta}
\def\D{\Delta}
\def\eps{\epsilon}
\def\g{\gamma}
\def\G{\Gamma}
\def\grad{\nabla}
\def\h{ {1\over 2}  }
\def\hc{\hat{c}}
\def\hd{\hat{d}}
\def\hg{\hat{g}}
\def\hp{ {+{1\over 2}}  }
\def\hm{ {-{1\over 2}}  }
\def\k{\kappa}
\def\l{\lambda}
\def\L{\Lambda}
\def\lg{\langle}
\def\m{\mu}
\def\n{\nu}
\def\o{\over}
\def\O{\Omega}
\def\p{\phi}
\def\pa{\partial}
\def\pr{\prime}
\def\ra{\rightarrow}
\def\rh{\rho}
\def\rg{\rangle}
\def\s{\sigma}
\def\t{\tau}
\def\th{\theta}
\def\ti{\tilde}
\def\wti{\widetilde}
\def\inte{\int dx }
\def\xb{\bar{x}}
\def\yb{\bar{y}}
%%                     common physics symbols
\def\tr{\mathop{\rm tr}}
\def\Tr{\mathop{\rm Tr}}
\def\partder#1#2{{\partial #1\over\partial #2}}
\def\ds{{\cal D}_s}
\def\wtwo{{\wti W}_2}
%%                    macros for Lie algebras
\def\lie{{\cal G}}
\def\alie{{\widehat \lie}}
\def\dlie{{\cal G}^{\ast}}
\def\elie{{\widetilde \lie}}
\def\edlie{{\elie}^{\ast}}
\def\hlie{{\cal H}}
\def\wlie{{\widetilde \lie}}
%%       fake blackboard bold macros for reals, complex, etc.
\def\rlx{\relax\leavevmode}
\def\inbar{\vrule height1.5ex width.4pt depth0pt}
\def\IZ{\rlx\hbox{\sf Z\kern-.4em Z}}
\def\IR{\rlx\hbox{\rm I\kern-.18em R}}
\def\IC{\rlx\hbox{\,$\inbar\kern-.3em{\rm C}$}}
\def\one{\hbox{{1}\kern-.25em\hbox{l}}}

\begin{titlepage}

December, 1991 \hfill{IFT-P.45/91}
\vskip .6in

\begin{center}
{\large {\bf Supersymmetric Construction of W-Algebras from Super Toda and
WZNW Theories}}\footnotemark \footnotetext{Work partially supported by CNPq}
\end{center}

\normalsize
\vskip .4in

\begin{center}

{L. A. Ferreira\footnotemark
\footnotetext{e-mail: 47553::LAF},
J. F. Gomes\footnotemark
\footnotetext{e-mail: 47553::JFG},
R. M. Ricotta\footnotemark
\footnotetext{e-mail: 47553::REGINA}
and A. H. Zimerman}
\par \vskip .1in \noindent
Instituto de F\'{\i}sica Te\'{o}rica-UNESP\\
Rua Pamplona 145\\
01405 S\~{a}o Paulo, Brazil
\par \vskip .3in

\end{center}

\begin{center}
{\large {\bf ABSTRACT}}\\
\end{center}
\par \vskip .3in \noindent

A systematic construction of super W-algebras in terms of the WZNW model based
on a super Lie algebra is presented.  These are shown to be the symmetry
structure of the super Toda models, which can be obtained from the WZNW theory
by Hamiltonian reduction.  A classification, according to the conformal spin
defined by an improved energy-momentum tensor, is discussed in general terms
for
all super Lie algebras whose simple roots are fermionic.  A detailed discussion
employing the Dirac bracket structure and an explicit construction of
W-algebras for the cases of $OSP(1,2)$, $OSP(2,2)$, $OSP(3,2)$ and $D(2,1\vert
\a)$ are given.  The N=1 and N=2 super conformal algebras are discussed in the
pertinent cases.
\end{titlepage}

\section{Introduction}

There has been an increasing demand in the study of the symmetry structures of
two dimensional conformal field theories. One of the interesting developments
in this direction was the introduction of extensions of the Virasoro algebra by
higher spin generators constituting the so called W-algebra \cite{zamo}. These
are not Lie algebras and have been shown to be the symmetries of a variety of
models \cite{bilal1,bakas,pope}. The geometrical and algebraic meaning of such
transformations are not well understood yet, but some results have already been
obtained in this direction \cite{bilal2,gervais}.

The W-algebras have been extensively studied in the context of the Toda models.
These can be cast in three different classes according to the algebraic
structure. There are the Conformal Toda models (CT) associated to finite Lie
algebras which are conformally invariant and which the simplest example is the
Liouville model. The Affine Toda models (AT) associated to the Kac-Moody
algebras with vanishing central extension (loop algebras) and which the
simplest
example is the Sinh-Gordon model. These are not conformally invariant but have
been shown to possess infinite number of conserved charges in involution
\cite{olive}. Finally there are the recently proposed Conformal Affine Toda
models (CAT) \cite{bonora,afgz} which are associated to the full Kac-Moody
algebra and constitute a conformal extension of the AT models.

The W structure appears in a very elegant way in the CT models through a
Hamiltonian reduction procedure. These models have been shown to be
constrained WZNW theories \cite{balog}. In order to preserve the conformal
invariance the energy momentum tensor has to be modified to commute with the
constraints. As a consequence,  the conformal spins of the currents are
changed giving rise to higher spin generators. The remaining Kac-Moody
currents under the reduction, which are the symmetries of the CT models, become
the generators of the W-algebra. Another framework for constructing the
W-generators for the CT models, involving differential operators, is based on
the work of Drinfeld-Sokolov \cite{ds,bilal1}. The CAT models have also been
shown to be obtained by a Hamiltonian reduction procedure from a WZNW type
model associated to a two loop Kac-Moody algebra \cite{afgz,schw}.
The W structure of
such models is not described just by the remaining Kac-Moody currents like in
\cite{balog}. There is an infinite number of W-generators and a method for
generating them was proposed in \cite{acfgz}. The AT models, being not
conformally invariant, were not studied along these lines. But an interesting
point to be explored in this context is the connection of the W symmetries of
the CAT models and the integrability structures of the AT models via a
breakdown of the conformal symmetry (see for instance \ct{break}).

The supersymmetric version of the CT models have also received a lot of
attention. They are superconformal models and constitute a natural place to
study the role of supersymmetry in the structures discussed above.  In
fact, several aspects of  the super W-algebras for such models have
been studied \cite{Hollowood}. A construction of $N=1$ and $N=2$ super
conformal algebra in terms of the fields of super conformal Toda models
(SCT) was proposed in \cite{nohara}. Their algebra was derived from the Poisson
brackets obtained from the SCT action. Recently, the SCT models have been shown
to be constrained super WZNW models associated to super Lie algebras whose
simple roots can be chosen to be all fermionic \cite{Inami1,olsha,Izawa}. Such
result paves the way to explore the symmetries of SCT models on the lines of
ref. \cite{balog}. Indeed, the spin of the W generators of those models were
calculated in ref. \cite{Sorba} using the methods of ref. \cite{balog}.

In this paper we propose a systematic construction of the super W-algebra for
the super conformal Toda models. Our starting point is a conformally invariant
WZNW model underlined by a super Lie algebra ${\cal G}$ whose simple roots are
all fermionic. Such theory, however is not supersymmetric because the number of
bosonic and fermionic generators in ${\cal G}$ do not match. In
section 2 we review the Hamiltionian reduction procedure which constrains our
model in a conformally invariant manner in order to obtain the SCT model,
which is supersymmetric. In
section 3 we give a general method to find the remaining super Kac-Moody
currents and their corresponding conformal spins. It is shown that both, the
bosonic and fermionic currents decompose each into rank ${\cal G}$ multiplets
of
a special $SL(2)$ subalgebra of ${\cal G}$. It is also argued in general terms
that after constraining and gauge fixing there is a single remaining current
associated to the highest weight of each $SL(2)$ multiplet. That constitute a
generalization to super Lie algebra of the analysis of ref. \cite{balog}.
Next in section 4 we analyze all cases where the simple roots are fermionic
and determine the conformal spin of each remaining current. In section 5
we present the super W-algebra for some examples by calculating explicitly the
Dirac brackets for the remaining currents. The examples discussed are: {\it
a)} $OSP(1,2)$ which possesses a $N=1$ superconformal algebra; {\it b)}
$OSP(2,2)$ which presents $N=2$ superconformal algebra; {\it c)}
$OSP(3,2)$ with an extension by a spin $5/2$ of the $N=1$ superconformal
algebra and finally {\it d)} the case of the exceptional super Lie algebra
$D(2,1 \mid \alpha )$ ($\alpha \neq 0,-1$) which presents three
non-commuting Virasoro subalgebras plus two generators of
spin $3/2$ and one of spin $5/2$. It is shown that such super W-algebra
possesses two non commuting $N=1$ superconformal subalgebras.

\section{The reduction of super WZNW theory}
We consider a WZNW theory based on a field $g(x)$ which takes value on a
connected real Lie supergroup $G$. Such theory possesses many of the properties
of the ordinary WZNW like conformal invariance and left and right Kac-Moody
symmetries. The equations of motion are given by
\be
\label{eq. motion}
\pa_{-} J_{R} = 0 \; ; \; \; \; \; \pa_{+} J_{L} = 0
\ee
where $J_{R}$ and $J_{L}$ are respectively the left and right Kac-Moody
currents
\be
\label{currents}
J_{R} =  k g^{-1} \pa_{+} g ; \; \; \; \; J_{L} =  - k  \pa_{-} g g^{-1}
\ee
where $k$ is the central term of the KM algebra.
In order to obtain the Super-Toda theories as reduced models from the above
WZNW models, the  superalgebra $\cal G$ of $G$ must be a Basic Lie superalgebra
which possess purely odd simple root system, \cite{Inami1,olsha}. These are
$sl(
   n+1\mid n)$
, $OSP(2n-1\mid 2n)$, $OSP(2n\mid 2n)$, $OSP(2n+1\mid 2n)$, $OSP(2n+2\mid
2n)$ and  $D(2,1\vert \a)$ with $n\geq 1$ and $\a \neq 0, -1$. The
supercommutation relations for $\cal G$, in the Chevalley basis, can be
written as
\br
\label{comrel}
\begin{array}{ll}
\mbox{$\lb H_a , H_b \rb  =  0 \; ;$} & \mbox{$\lb H_a , E_{\a_b} \rb  =
K_{ab} E_{\a_b} $} \\
\mbox{$\lb E_{\a_a} , E_{-\a_b} \rb  =  \d_{ab}H_a \; ;$} & \mbox{$\lb
H_a , E_{-\a_b} \rb  =  - K_{ab} E_{-\a_b}$}
\end{array}
\er
where $H_a$ are the generators of the Cartan subalgebra, $E_{\a_a}$ and
$E_{-\a_a}$ are the odd  operators associated to the positive and negative
simple roots respectively, $K_{ab}$ is the Cartan matrix of $\cal G$ (which can
be chosen real and symmetric), $\a_a$ are the simple roots of $\cal G$ (all
assumed to be odd roots) and $a,b = 1,2,...r$, where $r$ is the rank of $\cal
G$
   .
The superalgebra $\cal G$ possesses an invariant nondegenerate supersymmetric
bilinear form given by  \br
\label{bilinear form}
STr( H_a  H_b ) = K_{ab}; \; \; \; STr( E_{\a_a} E_{-\a_b}) = \d_{ab} \nn
\er
\br
STr( H_a  E_{\a_b}) = STr( H_a  E_{-\a_b} ) = STr( E_{\a_a}  E_{\a_b}) =
STr( E_{-\a_a} E_{-\a_b}) = 0
\er
The bilinear form for the remaining step operators $E_{\pm \a}$ is such that
$STr(H_a E_{\pm \a})=0$, and $STr(E_{\a} E_{\b})=0$ for $\a + \b \neq 0$. We
introduce the components of the KM currents
\be
\label{current components}
J_{R}(T) \equiv k STr(T g^{-1} \pa_{+} g ) \; ; \; \; \; J_{L}(T) \equiv - k
STr(T \pa_{-} g g^{-1})
\ee
Under the Poisson bracket each chiral component generates a copy of the
(super) KM algebra, and currents of different chiralities commute among
themselves.
The  energy-momentum tensor is of the Sugawara form,
\be
T(x) = {1\over{2k}}\eta ^{ij}J_i(x) J_j(x)
\label{EM}
\ee
where $\eta ^{ij}$ is the inverse of the Killing form, $\eta _{ij} =
STr(T_{i}T_{j})$, where $ T_{i}'s$ constitute a basis of $\lie$  and $J$
stand for either $J_L$ or $J_R$.  All currents have conformal
spin one with respect to $T(x)$, i.e.,
\be
\lb T(x),J_i(y) \rb = J_i(y)\d '(x-y) - J'_i(y)\d (x-y)
\ee
 The Super-Toda theories are obtained from
the WZNW model by a Hamiltonian reduction procedure where the following
constraints are imposed on the KM currents \cite{Inami1}
\br
\label{constraints}
\begin{array}{ll}
\mbox{$J_{L}(E_{\a_a})  =  0 \; ;$} &  \mbox{$J_{R}(E_{-\a_a})  =  0 $}\\
\mbox{$J_{L}(E_{\a_{ab}})  =  \mu^{-}_{(ab)} (1 + \d_{ab}) K_{ab} \; ;$} &
\mbox{$J_{R}(E_{-\a_{ab}})  =  -\mu^{+}_{(ab)}(1 + \d_{ab}) K_{ab} $} \\
\mbox{$J_{L}(E_{\a})  =  0 \; ;$} &  \mbox{$J_{R}(E_{-\a})  =  0$}
\end{array}
\er
where  $E_{\a_{ab}}$ and $E_{-\a_{ab}}$ are even simple root step operators
defined by
\be
\label{even simple roots}
E_{\a_{ab}} \equiv \lb E_{\a_a} , E_{\a_b} \rb ; \; \; \; \; E_{-\a_{ab}}
\equiv \lb E_{-\a_a} , E_{-\a_b} \rb
\ee
and $E_{\a}$ and $E_{-\a}$ are respectively the positive and negative remaining
step operators of $\cal G$ in the Chevalley basis. Using the super Jacobi
identity and the invariance of the bilinear form (\ref{bilinear form}) one gets
\be
\label{traco even simple}
STr( E_{\a_{ab}} E_{-\a_{cd}})  = - (\d_{ac} \d_{bd} + \d_{ad} \d_{bc})
K_{cd}
\ee
 The roots $\a_{ab} = \a_{a} + \a_{b}$, associated
to the step operators $E_{\a_{ab}}$, are the simple roots of the even
subalgebra
${\cal G}_0$, where ${\cal G} = {\cal G}_0 + {\cal G}_1$ is the ${\IZ}_2$
grading of $\cal G$. The reason is that $\a_{ab}$ can not be written as the
sum of two positive even roots of $\cal G$. Notice that ${\cal G}_0$ may not be
a simple ordinary Lie algebra, and its set of simple roots is the union of the
sets of simple roots of each one of its simple components.

The equations of motion of the Super-Toda theories are obtained from
(\ref{eq. motion}) by parametrizing the group elements close to the identity
by a Gauss type decomposition
\be
\label{gauss}
g = {\cal NAM}
\ee
where
\br
\label{gauss components}
\cn (x_{+},x_{-}) = \exp {\cal F}_{+}(x_{+},x_{-}) \; ; &\; \; & \cm
(x_{+},x_{-}) = \exp {\cal F}_{-}(x_{+},x_{-}) \; ; \nn \\
\ca (x_{+},x_{-}) & = & \exp (\phi_a (x_{+},x_{-}) H_a )
\er
and
\be
\label{def fermionic fields}
\cf_{+} = \sum_{a=1}^{rank{\cal G}} \zeta_a (x_{+},x_{-}) E_{\a_a} +
\tilde{\cf}_{+}\; ; \; \; \cf_{-} = \sum_{a=1}^{rank{\cal G}} \xi_a
(x_{+},x_{-}) E_{-\a_a} + \tilde{\cf}_{-}
\ee
where $\tilde{\cf}_{+}$ and $\tilde{\cf}_{-}$ are real linear combinations of
the positive and negative non simple roots of $\cal G$ respectively, $\phi_a $
are bosonic fields and $\zeta_a$ and $\xi_a$ are grassmannian fields.

We introduce
\br
\label{def K}
K_R  \equiv  \cm g^{-1} \pa_{+} g \cm^{-1} & = & \ca^{-1} \cn^{-1}
\pa_{+} \cn \ca + \ca^{-1} \pa_{+} \ca + \pa_{+} \cm \cm^{-1} \nn \\
K_L  \equiv  \cn^{-1} \pa_{-} g g^{-1} \cn & = & \cn^{-1} \pa_{-} \cn +
\pa_{-} \ca \ca^{-1} + \ca \pa_{-} \cm \cm^{-1} \ca^{-1}
\er
The equations of motion (\ref{eq. motion}) can then be written as
\br
\label{eqmov}
\pa_{-} K_R & = & - \lb K_R , \pa_{-} \cm \cm^{-1} \rb \\
\label{eqmovL}
\pa_{+} K_L & = & \lb K_L , \cn^{-1} \pa_{+} \cn \rb
\er
After the constraints (\ref{constraints}) are imposed, the currents take the
for
   m
\br
\label{constrained currents R}
J_R = k g^{-1} \pa_{+} g & = & \sum_{(ab)} \mu_{(ab)}^{+} E_{\a_{ab}} + j_R  \\
\label{constrained currents L}
J_L = - k \pa_{-} g g^{-1} & = & \sum_{(ab)} \mu_{(ab)}^{-} E_{-\a_{ab}} + j_L
\er
where $j_R$ ($j_L$) is a linear combination of Cartan subalgebra generators and
of negative (positive) root step operators. From (\ref{even simple roots}) and
the invariance of the bilinear form $STr$ it follows that
\br
\lb E_{\a_{ab}} ,E_{-\a_c} \rb = - K_{ab}( \d_{bc} E_{\a_a} + \d_{ac}E_{\a_b})
\nn \\
\lb E_{-\a_{ab}} ,E_{\a_c} \rb =  K_{ab}( \d_{bc} E_{-\a_a} + \d_{ac}E_{-\a_b})
\er
{}From (\ref{def K}), (\ref{constrained currents R}) and
(\ref{constrained currents L}) we conclude that, after the constraints are
imposed
\br
\label{positive KR}
\(\ca^{-1} \cn^{-1} \pa_{+} \cn \ca \)_{constr.} & = &  \sum_{(ab)}
\mu_{(ab)}^{+} \( E_{\a_{ab}} + K_{ab} \( \xi_b E_{\a_a} + \xi_a E_{\a_b} \) \)
\nn \\
\label{negative KL}
\( \ca \pa_{-} \cm \cm^{-1} \ca^{-1} \)_{constr.} & = & \sum_{(ab)}
\mu_{(ab)}^{-} \( E_{-\a_{ab}} + K_{ab} \( \zeta_b E_{-\a_a} + \zeta_a
E_{-\a_b}
\) \)
\er
Further, after a conjugation by the abelian subgroup $\ca$, we find
\br
\label{positive KR1}
\(\cn^{-1} \pa_{+} \cn \)_{constr.}  =   \sum_{(ab)}
\mu_{(ab)}^{+} \( e^{(K_{ac}+K_{bc})\phi _{c}}E_{\a_{ab}} + K_{ab} \( \xi_b
e^{K_{ac}\phi _{c}}E_{\a_a} + \xi_a e^{K_{bc}\phi _{c}}E_{\a_b} \) \) \nn \\
\label{negative KL1}
\( \pa_{-} \cm \cm^{-1} \)_{constr.} = \sum_{(ab)}
\mu_{(ab)}^{-} \(e^{(K_{ac}+K_{bc})\phi _{c}} E_{-\a_{ab}} + K_{ab} \( \zeta_b
e^{K_{ac}\phi _{c}}E_{-\a_a} + \zeta_ae^{K_{bc}\phi _{c}} E_{-\a_b} \) \)
\er
Now using (\ref{negative KL}) and (\ref{negative KL1}) in (\ref{eqmov}) and
(\ref{eqmovL}), we  find the equations of motion for the fields $\phi_a$, $\;
\xi_a$ and $\; \zeta_a$.  The equation for $\phi_a$ correspond to the
coefficients of the Cartan subalgebra in either (\ref{eqmov}) or
(\ref{eqmovL}), whilst the coefficient of the positive and negative simple
root step operator of (\ref{eqmovL}) and (\ref{eqmov}), respectively, yields
the equations of motion for $\xi_a$ and $\; \zeta_a$.   They are
\br
\pa _{+}\pa _{-}\phi _{a} & = & 4\sum_b K_{a,b} \mu _{ab}^{+}\mu_{ab}^{-}
e^{(K_{ac}+K_{bc})\phi _{c}} -4 \sum_{b,d} \mu
_{ab}^{+}\mu_{ad}^{-}K_{ab}K_{ad}
\xi _{b}\zeta _{d}e^{K_{ac}\phi _{c}} \nonumber \\
\pa _{-}\xi _{a} & = & 2\sum_b \mu _{ab}^{-}K_{ab}\zeta
_{b}e^{K_{ac}\phi_{c}} \\
\pa _{+}\zeta _{a} & = & 2\sum_b \mu _{ab}^{+}K_{ab}\xi
_{b}e^{K_{ac}\phi_{c}}. \nonumber
\er
These are the equations of motion of the super conformal Toda models.
\section{The Super $W$-algebra}
The symmetries of the WZNW theory are given by the left and right KM currents
(\ref{current components}). The symmetries of the Super-Toda theories are
described by the currents which remain after the constraints
(\ref{constraints})
and the corresponding gauge fixings are imposed. The algebra of these remaining
currents, under the Dirac bracket, is not a subalgebra of the KM algebra. In
fact, it is not even a Lie algebra. Their algebra constitute what is called a
super $W$-algebra. We now describe a method to obtain the $W$ generators and
their algebra. It is a generalization to the case of superalgebras, of the
method described in \cite{ds,balog}.
Our discussion applies equally well to both chiralities. So, we will restrict
ourselves to the right sector.  In order for the constraints
(\ref{constraints}) to preserve conformal invariance the energy-momentum
tensor (\ref{EM}) needs to be modified to
\be
L(x) = T(x) + 2\partial _{x}J_{\tilde \d .H}(x)
\label{impEM}
\ee
where $J_{\tilde \d .H}(x) = STr(J(x)\tilde \d .H)$ and $\tilde \d =
{1\over{2}}\sum_{\a >0} {\a \over {\a ^2}}$, one half of the sum of the
even coroots of $\lie_0$.  With respect to the improved energy-momentum tensor
$L(x)$ the conformal spins of the currents $J_i$ are changed.  In particular,
$J_R (E_{-\a_{ab}})$ become scalars.  Those corresponding to the Cartan
subalgebra remain unchanged whilst those corresponding to positive root step
operators are increased as
\be
\lb L(x),J(E_{\a}(y) \rb =(1+h(\a)) J(E_{\a})(y)\d '(x-y) - J'^a(y)\d (x-y)
\label{newspin}
\ee
where $h(\a) = 2\tilde \d . \a $ is called the height of the root $\a $.
After the constraints (\ref{constraints}) are imposed, the current $J_{R}$
takes the form (\ref{constrained currents R}). The constant operator
\be
\label{iplus}
I_{+} = \sum_{(ab)} \mu^{+}_{(ab)} E_{\a_{ab}}
\ee
appearing on the r.h.s of (\ref{constrained currents R}) plays a crucial role
in what follows. It belongs to a special $Sl(2)$ subalgebra $\cal S$ of ${\cal
G}_0$
\be
\label{su2}
\lb T_3 , I_{\pm} \rb = \pm I_{\pm} ; \; \; \; \; \lb I_{+}, I_{-} \rb = 2 T_3
\ee
where
\be
\label{su2 generators}
T_3 = 2 {\tilde{\d}}. H ; \; \; \; \; I_{-} = \sum_{(ab)} w_{(ab)} E_{-\a_{ab}}
\ee
where $w_{(ab)}$ are determined by imposing (\ref{su2}). Since $T_3$ is an
element of the Cartan subalgebra, the step operators of $\cal G$ are its
eigenstates
\be
\lb T_3 , E_s \rb = h(s) E_s
\label{height}
\ee
 The simple roots
$\a_{ab}$ of ${\cal G}_0$ have unit height. From the super Jacobi
identity it follows that if $s$, $s'$ and $s + s'$ are roots then $h(s + s') =
h(s) + h(s')$. Therefore the simple roots of $\cal G$ must satisfy
$h(\a_a) + h(\a_b) = 1$ for any pair $\a_a , \a_b$. The adjoint
representation of $\cal G$ can be decomposed into irreducible representations
of  $\cal S$. Since these are finite, it follows that the eigenvalues of
$T_3$ must be integers or half integers. Therefore one concludes that
\be
h(\a_a) = \h
\ee
for any simple root $\a_a$ of $\cal G$.
The decomposition of the adjoint representation of $\cal G$ into
multiplets of $\cal S$ is very useful in what follows. Since $\cal S$
contains only even generators its multiplets will be constituted of only
even or only odd generators of $\cal G$. In fact the adjoint of ${\cal
G}_0$ itself decomposes into multiplets of $\cal S$. The weights of $\cal
S$ appearing in it are all integers (zero and the heights of the even roots).
The generators of ${\cal G}_1$ define an even dimensional representation of
${\cal G}_0$. Indeed the odd roots of $\cal G$ are the weights of this
representation. Such representation of ${\cal G}_0$ will also break into
multiplets of $\cal S$. The weights of $\cal S$ appearing in it are all half
integers. In an irreducible representation of $\cal S$ the eigenvalues of $T_3$
are not degenerate and so it follows that the Cartan subalgebra generators of
$\cal G$ must belong to different multiplets. Since multiplets with integer
spin necessarily contains the zero weight, the number of $\cal S$-multiplets in
${\cal G}_0$ is exactly the rank of $\cal G$. For the same reason the simple
roots of $\cal G$ must belong to distinct multiplets. Since any positive
(negative) odd root necessarily has height greater than $\h$ (smaller than or
equal to $-\h$) they must belong to one of the multiplets where the simple root
step operators are. Therefore the representation of ${\cal G}_0$ on ${\cal
G}_1$ also decomposes into exactly rank $\cal G$ $\cal S$-multiplets. In
ref.\cite{Sorba} the adjoint representation of $\lie $ was decomposed into
super multiplets of a special $OSP(1,2)$ subalgebra.

We now discuss the choice of gauge fixing. We want the Poisson bracket
of a constraint with its respective gauge fixing to be proportional to
the currents $J_{R}(E_{-\a_{ab}}) $ which are set to constants in
(\ref{constraints}). Therefore the gauge fixing of the constraints
\be
\varphi_{(ab)} = J_{R}(E_{-\a_{ab}}) + \mu^{+}_{(ab)}(1 + \d_{ab})
K_{ab}
\ee
can be taken to be the Cartan subalgebra generators $J_{R}(H_a)$.
Notice that if ${\cal G}_0$ has $U(1)$ factors the number of even simple
roots $\a_{ab}$ is smaller than the rank of $\cal G$. Therefore one does
not have all the Cartan subalgebra generators as gauge fixing. The gauge
fixing of the constraint $J_{R}(E_{-\a}) $, where $\a$ is an even positive
non simple root, can be taken to be $ J_{R}(E_{\b}) $ where $\b$ is a
positive even root such that $\lb E_{\b} , E_{-\a} \rb$ is proportional to an
even  negative simple root step operator $E_{-\a_{ab}}$. This means that $h(\a
)
- h(\b ) = 1$. By doing this one can convince oneself that there will always
be a remaining current (not used as gauge fixing) of height $j$
whenever there is a $\cal S$-multiplet  with highest weight $j$. Therefore the
number of remaining currents with integer height is equal to the number of
$\cal S$-multiplets in the adjoint of ${\cal G}_0$ which is equal to rank $\cal
G$. This is in fact the result discussed in \cite{balog}.

We can now apply the same procedure to choose the gauge fixing of the
constraints associated to the odd roots. The gauge fixing of
$J_{R}(E_{-\a_a})$,
$a = 1,2,...r$ can be taken to be themselves, since their Poisson brackets are
proportional to  $J_{R}(E_{-\a_{ab}}) $ which are set to constants in
(\ref{constraints}). The gauge fixing of the constraints $J_{R}(E_{-\sigma})$,
where $\sigma$ is an odd positive non simple root, can be chosen to be
$J_{R}(E_{\sigma '})$ where $\sigma '$ is a positive odd root such that
$\lb E_{-\sigma} , E_{\sigma '} \rb$ is to proportional to a negative even
simpl
   e
root operator $E_{-\a_{ab}}$. Again $h(\sigma ) - h(\sigma ') = 1$. By the same
reasoning as above one concludes that the number of remaining currents with
half
integer height is equal to the number of $\cal S$-multiplets in ${\cal G}_1$
which is also equal to the rank of $\cal G$.

Summarizing, we have shown that
\begin{enumerate}
\item The adjoint representation of ${\cal G}_0$ decomposes into rank $\cal
G$ multiplets of the subalgebra $\cal S$ (\ref{su2}).
\item The representation of ${\cal G}_0$ in ${\cal G}_1$ also decomposes
into rank $\cal G$ $\cal S$-multiplets.
\item After the gauge fixing there is a one to one correspondence between
remaining currents and $\cal S$-multiplets in the adjoint of $\cal G$.
\item The number of remaining currents with integer and half integer height is
the same and equal to the rank of $\cal G$.
\item The conformal spin, w.r.t. $L(x)$, of the remaining current associated
to the $\cal S$-multiplet with highest weight $ j$ is $(j+1)$.
\end{enumerate}

We now show that the remaining currents can be written in terms of the
Super-Toda fields and its derivatives only, showing that they are indeed
the symmetries of the WZNW theory which survives the reduction
procedure. From (\ref{def K}) and (\ref{positive KR}) one gets
\be
\label{constr. JR}
J^{constr.}_{R} = k \( g^{-1} \pa_{+} g \)_{constr.} = \cm^{-1} \( I_{+} +
I_{\h
   }
+ \pa_{+} \Phi \) \cm + \cm^{-1} \pa_{+} \cm = I_{+} + j_{R}
\ee
where $\pa_{+} \Phi = \ca^{-1} \pa_{+} \ca = \sum_{a=1}^{rank {\cal G}}
\pa_{+}\phi_a H_a $, $I_{+}$ is defined in (\ref{iplus}), and
\be
I_{\h} \equiv \sum_{(ab)} \mu_{(ab)}^{+} K_{ab} \( \xi_b E_{\a_a} + \xi_a
E_{\a_b} \)
\ee
The fields of the Super-Toda theory are the parameters $\phi_a$ of the abelian
subgroup $\ca$, $\zeta_a$ and $\xi_a$ ($a=1,2,...$ rank $\cal G$). The fields
of the WZNW model which we want to eliminate from the remaining currents, and
which appear in (\ref{constr. JR}) are the parameters of the subgroup $\cm$
contained in $\tilde{\cf}_{-}$ (see (\ref{def fermionic fields})). We now show
that after the gauge fixing described above the remaining currents depend only
on the fields of the Super-Toda theory and therefore are symmetries of it.

The analysis is made simpler by grading the generators with eigenvalues of
$T_3$ defined in (\ref{su2 generators}). We write $\tilde{\cf}_{-} =
\sum_{s} \tilde{\cf}_{-}^{-s}$, and $(J_{R})^{constr.} = I_{+} +
\sum_{s'}J_{R}^{-s'}$, where
\be
\lb T_3 , \tilde{\cf}_{-}^{-s} \rb = - s \tilde{\cf}_{-}^{-s} \; ; \; \; \;
\lb T_3 , J_{R}^{-s'} \rb = - s' J_{R}^{-s'}
\ee
We then have
\br
J_R^1 & = & I_{+} \nn \\
J_R^{\h} & = & 0 \nn \\
J_R^0 & = & \lb I_{+} , \tilde{\cf}_{-}^{-1} \rb + \lb I_{\h} , \xi_a
E_{-\a_a}\rb + \pa_{+} \Phi + \h \lb \xi_a E_{-\a_a},\lb \xi_b E_{-\a_b}, I_{+}
\rb \rb \nn \\
J_R^{-\h} & = & \lb I_{+} , \tilde{\cf}_{-}^{-3/2} \rb + ...
\er and so on. Therefore
\be
J_R^{-s} =  \lb I_{+} , \tilde{\cf}_{-}^{-s -1} \rb +  X^{-s}
\ee
where $X^{-s}$ involves $\tilde{\cf}_{-}^{-s'}$ for $s'\leq s + \h$ only. We
can then write the fields appearing in $\tilde{\cf}_{-}$ in terms of the fields
of the Super-Toda theory by gauge fixing the currents recursively. For
instance, we set to zero a number of components of $J_R^0$ equal to the
dimension of the subspace generated by $\lb I_{+} , \tilde{\cf}_{-}^{-1} \rb $.
That is equal to rank $\cal G$ except for the cases where ${\cal G}_0$ has
$U(1)$ factors. We then have the fields in $\tilde{\cf}_{-}^{-1}$ written in
terms of $\phi_a$ and $\xi_a$. Analogously we set to zero a number of
components of $J_R^{-\h}$ equal to the dimension of the subspace generated by
$\lb I_{+} , \tilde{\cf}_{-}^{-3/2} \rb$ and eliminate the fields in
$\tilde{\cf}_{-}^{-3/2}$. Therefore one observes that the number of remaining
currents (not used as gauge fixing) of height $s$ is equal  the number of
components of $J_R^{-s}$ minus the dimension of the subspace spanned by $\lb
I_{+},\tilde{\cf}_{-}^{-s -1} \rb$. But this is exactly the number of
multiplets of $\cal S$ with highest weight $s$, showing that this is the same
gauge fixing discussed above. Since at the end of the process the fields in
$\tilde{\cf}_{-}$ will all be written in terms of the fields of the Super-Toda
theory, so will all the remaining currents. These currents will be generators
of symmetries of  the Super-Toda theory. As we have shown before their number
is twice the rank of $\cal G$. Half of them have integer height and the other
half have half integer height.

Notice that the remaining gauge symmetry, after the constraints
(\ref{constraints}) are imposed, is given by the subgroup generated by
negative non simple roots. Indeed, the form of the constrained current
(\ref{constrained currents R}) is left unchanged by the gauge transformation
\be
J_{R} \ra {\tilde {\cal M}}^{-1} J_{+} {\tilde {\cal M}} + {\tilde {\cal
M}}^{-1} \pa_{+} {\tilde {\cal M}}
\ee
where ${\tilde {\cal M}}$ is a exponentiation of a real linear combination of
the step operators corresponding to negative {\em non simple} roots. The
negative simple root step operators can not be included because their super
commutator  with $E_{\a_{ab}}$ would produce positive simple root step
operators.

\section{Higher Spin Generators for the Super Toda Model}

In this section we discuss in detail how the framework of section 3 can be
applied to the cases where $ \lie _1$ contain all the simple roots of $ \lie
$.  We only deal with the fermionic sector since the decomposition of the
adjoint representation of ${\lie _0}$ into multiplets of $Sl(2)$ has been
discussed by Balog et al \cite{balog}. The $ T_3 $ generator defined in
eq.(\ref
   {su2 generators}) is constructed
in terms of the roots of $ \lie _0$  and these in terms of a set of
unit length vectors, $\epsilon_i's$ and $\rho_a's$, where $\epsilon_i \cdot
 \epsilon_j = - \delta_{i,j}$,$\;\;$ $\rho_a \cdot \rho_b = \delta_{a,b}$.
Let us consider case by case all possibilities.

\nit 1.  $\lie = OSP(2l+1,2n), \;\;\;\;\;\;\; \lie _0 = SO(2l+1)\otimes SP(2n),
\;\;\;\;\; n,l=1,2,...  $

\nit The even and odd set of roots according to \cite{Frappat} are given
respectively as

\br
\Delta _0 = \lbrace \pm \epsilon _i \pm \epsilon _j, \pm \epsilon _i, \pm \rho
_a \pm \rho _b, \pm 2\rho _a, \;\;i,j=1,2,...l,\;\; a,b=1,2,...n \rbrace
\nonumber
\er
\be
\Delta _1 = \lbrace \pm \rho _a, \pm \epsilon _i \pm \rho _a \rbrace \nonumber
\ee
  The height $h(a)$ of a root $a$ is defined in
(\ref{height}) and is determined in terms of $\tilde \d = {1 \over{2}} \sum
_{\a
>0} {\a
\over {\a ^2}}$.  Let
\be
 \tilde \d _{SP(2n)}\; = \; {1 \over {4}}\sum _{a<b}^n
{(\rho _a - \rho _b)} + {1 \over {4}}\sum _{a<b}^n {(\rho _a +
\rho _b)} + {1 \over {4}}\sum _{a=1}^n {\rho _a } \; = \; {1
\over {2}}\sum _{a=1}^n (n-a+{1 \over {2}})\rho _a  \label{deltasp2n}
\ee
and
\be
\tilde \d _{SO(2l+1)} \; = \; - { 1 \over {4}}\sum _{i<j}^n
{(\epsilon _i - \epsilon _j)} - {1 \over {4}}\sum _{i<j}^n {(\epsilon _i +
\epsilon _j)} - {1 \over {2}}\sum _{i=1}^n {\epsilon _i }\; = \; - {1
\over {2}}\sum _{i=1}^l (n-i+1)\epsilon _i \label{deltasoimpar}
\ee
from where we derive
\be
h(\rho _a) = \tilde \d _{SP(2n)} \cdot \rho _a = n-a+{1\over {2}} \label{hrho}
\ee
\be
h(\epsilon _i) = \tilde \d _{SO(2l+1)} \cdot \epsilon _i = l-i+1
\label{hepsilon
   }
\ee
The number $N({{2j+1}\over 2})$ of odd roots of height ${{2j+1}\over {2}}$ can
be evaluated for the two cases of interest, i.e.

\nit $1.a)\;\;$ $l=n-1$, $\;\;$i.e. $\;\; \lie = OSP(2n-1,2n)$

\be
N({{2j+1}\over {2}}) = 2n-j-1
\ee

\nit $1.b)\;\;$ $l=n$, $\;\;$ i.e. $\;\;$ $\lie = OSP(2n+1,2n)$

\be
N({{2j+1}\over {2}}) = 2n-j
\ee
For both cases, when $j=0$, $N({1\over{2}})$ yields the number of $Sl(2)$
multiplets $\Delta _1$ is decomposed.  This agrees with the general argument of
the previous section to be rank $\lie $.  Further, since there is a remaining
current associated to the highest weight of each multiplet, the number of
W-generators of conformal spin ${{{2j+3} \over {2}} }$ is therefore found to be
equal to
\be
N({{2j+1}\over {2}}) - N({{2j+3}\over {2}})  =  1
\ee

\nit $2.)\;\;$ $\lie$ = $OSP(2l,2n)$, $\;\;\;\;\;\;\;\;$ $\lie _0$ =
$SO(2l)\otimes SP(2n)$
\br
\Delta _0 = \lbrace \pm \epsilon _i \pm \epsilon _j, \pm \rho_a \pm \rho _b,
\pm 2\rho _a, \;\;i,j=1,2,...l,\;\; a,b=1,2,...n \rbrace \nonumber
\er
\br
\Delta _1 = \lbrace \pm \epsilon _i \pm \rho _a \rbrace \nonumber
\er
Again the height of the roots of $OSP(2l,2n)$ is given by eq.(\ref{height})
where
\be
\tilde \d _{SO(2l)} \; = \; - {1 \over {2}}\sum _{i<j}^l {(\epsilon _i -
\epsilon _j) } - {1 \over {2}}\sum _{i<j}^l {(\epsilon _i +
\epsilon _j)} \; = \; - {1 \over {2}}\sum _{i=1}^l (l-i)\epsilon _i
 \label{deltasopar}
\ee
It then follows
\be
h(\epsilon _i) = \tilde \d _{SO(2l)} \cdot \epsilon _i = l-i
\ee
For the cases of interest, i.e.

\nit $2.a)\;\; $ $l=n$,$\;\;$ i.e. $\;\;$ $\lie = OSP(2n,2n)$, $n > 1$,
\be
 N({{2j+1}\over {2}}) = 2n-j
\ee
\nit $2.b)\;\; $ $l=n+1$,$\;\;$ i.e.$\;\;$ $\lie = OSP(2n+2,2n)$
\be
   N({{2j+1}\over {2}}) = 2n-j +1
\ee
\nit In all cases discussed so far there is a single W-generator of conformal
spin ${{2j+3}\over {2}}$.

\nit $3.)\;\;$ $\lie$ = $OSP(2,2)$, $\;\;\;\;\;\;\;$ $\lie _0$ = $SO(2)\otimes
SP(2)$
\br
\Delta _0 = \lbrace \pm 2\rho \rbrace \nonumber
\er
\br
\Delta _1 = \lbrace \pm \epsilon \pm \rho \rbrace  \nonumber
\er
The $T_3$ generator is given in terms of the fundamental weight of the bosonic
subalgebra $\lie _0$,
\be
T_3 = {1\over {2}}\rho.H
\label{rhoh}
\ee
The fermionic roots in $\Delta _1$  decomposes therefore into two doublets,
namely $\epsilon \pm \rho $ and $-\epsilon \pm \rho $ yielding two remaining
currents of conformal weight ${3 \over {2}}$, i.e.
\be
G_+ = J_{\epsilon + \rho}, \;\;\;\;\;\;\;\ G_- = J_{-\epsilon + \rho}
\ee
This two generators of spin ${3 \over {2}}$ give rise to a $N=2$
superconformal theory.  This fact shall be explicitly shown in the next
section using Dirac brackets.

\nit $4.)\;\;$ $\lie$ = $SU(n+1,n)$,$\;\;\;\;\;\;\;$ $\lie _0$ =
$SU(n+1)\otimes
SU(n)\otimes U(1)$
\br
\Delta _0 =\lbrace \pm (\epsilon _i - \epsilon _j), \pm (\rho _a -\rho _b),\;\;
i,j =1,2,...n+1,\;\; a,b=1,2,...n \rbrace \nonumber
\er
\br
\Delta _1 = \lbrace \pm (\epsilon _i - \rho _a )\rbrace \nonumber
\er
The height of the fermionic roots in $\Delta _1$ is determined from the
definition of $T_3$ in eq.(\ref{su2 generators}) where
\be
\tilde \delta _{su(l)} = {1\over {2}}\sum _{i<j}^{l+1} {(\s _i - \s _j)
\over {\vert  \s _i - \s _j \vert}^2} =
{1\over{4}}\sum _{i=1}^{l+1} (l-2i+1){\s _i \over {\vert \s _i \vert}^2}
\ee
where $\s_i$ stand for either $\epsilon _i$ or $\rho _a$.  Therefore
\be
h(\epsilon _i - \rho _a) = a-i+{1\over{2}}
\ee
The number of odd roots of height  ${{2j+1}\over {2}}$ is then given by
\be
N({{2j+1}\over {2}}) = 2n-2j
\ee
Hence the number of W-generators of conformal spin  ${{2j+3}\over {2}}$ is then
\be
N({{2j+1}\over {2}}) -  N({{2j+3}\over {2}})= 2.
\ee
This example shows that there are always $2$ $W$-generators of conformal spin
${{2j+1}\over {2}}$ and hence should be associated to a $N=2$  superconformal
theory.

\nit $5.)\;\;$ $\lie$ = $D(2,1|\a)$,$\;\;\;\;\;\;\;$ $\lie _0$ = $SU(2)
\otimes SU(2) \otimes SU(2)$, $\a \neq 0,-1$.  The bosonic and
fermionic roots are, respectively
\be
\Delta _0 = \lbrace \pm2\epsilon _i,\;\;\;i=1,2,3 \rbrace
\ee
\be
\Delta _1 = \lbrace \pm \epsilon _1 \pm \epsilon _2 \pm \epsilon _3 \rbrace
\ee
where $\epsilon _1^2 = -{\a + 1 \over 2}$, $\;\; \epsilon _2^2 = {1\over2}$
and $\epsilon _3^2 = {\a \over2}$.  The $T_3$ generator is then defined to be
\be
T_3 = 2\tilde \delta \cdot H =
\left ( -{\epsilon_1 \over \a +1} + \epsilon _2 + {\epsilon _3 \over
\a} \right )  \cdot H
\ee
The $8$ odd roots of $\lie$ therefore decomposes into 2 doublets and one
quadruplet  as follows
\br
h(\epsilon _1 + \epsilon _2 + \epsilon _3)= {3 \over{2}} \nonumber
\er
\br
h(\epsilon _1 + \epsilon _2 - \epsilon _3) =
h(\epsilon _1 - \epsilon _2 + \epsilon _3) =
h(- \epsilon _1 + \epsilon _2 + \epsilon _3) = {1 \over {2}} \nonumber
\er
\nit  Although there are two W-generators of conformal  weight ${3 \over
{2}}$ this model contains two non-commuting $N=1$ superconformal
systems, whose structure is displayed in Section 5.

\section{Examples}

In this section we discuss in detail how the structure described in the
previous sections can be illustrated to construct Super W-algebras.  In
particular, this presents a construction of representations of N=1 and N=2
super-conformal algebras in terms of the super Toda fields. Our notation for
the
super Kac-Moody algebra is as follows,

\br
\label{super algebra}
[J_{H_i}(x), J_{H_j}(y)] = k STr (H_i H_j) \d ' (x-y) \nonumber
\er
\br
[J_{H_i}(x), J_{\pm \a} (y)] = \pm \a \cdot \epsilon_i J_{\pm \a} (y) \d (x-y)
\er
\br
[J_{\a} (x), J_{\b}(y)] = \left\{ \begin{array}{ll}
\varepsilon (\a , \b)J_{\a +\b}(x) \d(x-y) &   \mbox{if $\a +\b$ is a root} \\
B_{\a} \a_i J_{H_i}\delta (x-y) + kB_{\a} \d ' (x-y)
& \mbox {if $\a +\b = 0 $} \\
0 & \mbox {otherwise}
\end{array}
\right.  \nonumber
\er
where $B_{\a} =  STr(E_{\a} E_{-\a})$, $\a = \sum_{i=1}^{rank {\cal G}} \a_i
\epsilon_i$ and $\epsilon_i's$ constitute a basis for the root space. Much of
the calculation  in this section can be performed using a computer program for
algebraic manipulation.

\subsection{{\cal G} = OSP(1,2) and the N = 1 Super Conformal Theory}

Let us consider the super Lie algebra $OSP(1,2)$ where the even and odd
roots are, respectively  $\pm 2\rho$ and $\pm \rho$, (${\rho}^2 = 1$).
The Lie algebra is
\br
[H, E_{\pm 2\rho} ] = {\pm} 2 E_{\pm 2\rho} \;; \; \; \; [H, E_{\pm \rho} ] =
\pm E_{\pm \rho}
\er
\br
[E_{+ 2\rho}, E_{-2\rho}] = 2B_{2\rho}H \;;\;\;\;
\{ E_{+\rho}, E_{-\rho}\}= B_{\rho}H\;;\nonumber
\er
\br
\{E_{\pm \rho}, E_{\pm \rho}\} = \varepsilon (\pm \rho , \pm \rho )
E_{\pm 2\rho}\;;\;\;\;
[E_{\pm 2\rho}, E_{\mp \rho}] = \varepsilon (\pm 2 \rho ,
\mp \rho) E_{\pm \rho} \nonumber
\er
\br
[E_{\pm 2\rho}, E_{\pm \rho}] = 0 \nonumber
\er
Since we are already dealing with a conformal invariant field theory we
shall only consider one chirality, say $ J_R = J $. Following the Gauss
decomposition,(\ref{gauss})-(\ref{def fermionic fields}), we shall describe
as bosonic fields are $\phi(x)$, $f(x)$ and $g(x)$ and the
fermionic fields as $\zeta(x)$  and $\xi(x)$.
{}From  (\ref{gauss components}) we write,
\be
\cf_{+} = \zeta (x) E_{\rho} + f(x) E_{2\rho}  \;\;\;;
\cf_{-} = \xi (x) E_{-\rho} + g(x) E_{-2\rho}
\ee
and find the components of the  current defined in (\ref{current components})
to
be (normalizing $Str H^2 = 1$),
\br
{1\over k}J_{-2\rho} = B_{2\rho}
(\pa_+f- {\varepsilon (\rho , \rho )\over 2} \zeta  \pa_+{\zeta}) e^{-2\phi}
\er
\br
{1\over k}J_{H} ={2 \over k}g J_{-2\rho} - B_{\rho} \xi \pa_+ {\zeta}
e^{-\phi} +  \pa_+ \phi \nonumber
\er
\br
{1\over k}J_{2\rho} = B_{2\rho}\left ( -{2\over k}g^2 J_{-2\rho} + 2B_{\rho}
g \xi \pa_+{\zeta} e^{-\phi} -2 g \pa_+{\phi} + \pa_+
g - {1\over 2} \varepsilon (\rho , \rho ) \xi \pa_+ \xi \right ) \nonumber
\er
\br
{1\over k}J_{\rho} = B_{\rho}\left ( {2\over k}g \xi J_{-2\rho} -
\varepsilon (2\rho, -\rho)
g \pa_+ \zeta e^{-\phi} +  \xi \pa_+ \phi  -  \pa_+ \xi \right ) \nonumber \er
\br
{1\over k}J_{-\rho} =  B_{\rho}\left ( {\varepsilon (2\rho, -\rho) \over
k B_{2\rho}} \xi
J_{-2\rho} + \pa_+ \zeta e^{-\phi} \right )\nonumber
\er

All currents have conformal weight one, with respect to the
Sugawara energy momentum tensor, (\ref{EM}). However,
with respect to the modified energy-momentum tensor,
\be
L(x) = T(x) + \pa_{x} J_{H_{\rho}}
\ee
the current $J_{-2\rho} $ becomes scalar and can be set to a constant
$\lambda$ without breaking the conformal symmetry.
Also following (\ref{constraints}) the current associated to the negative
fermionic root $J_{-\rho} $ should be set to vanish. These two constraints
imply further subsidiary conditions (gauge fixings) to be imposed.
In order to evaluate the Dirac brackets of the remaining currents,
we need to invert the matrix
\be
\label{matriz delta}
\Delta_{i,j}(x, y) = {\left\{ \psi_i(x), \psi_j(y)
\right\}}_{PB}\|_{constrained
   }
\;; \;\;
i, j = 1, 2, 3.
\ee
where the constraints and gauge fixings are
\be
\label{gauge fixing}
\psi_1 = J_{-2\rho} - \lambda  \;;\;\;\; \psi_2 = J_H \;;\;\;\;
\psi_3 = J_{-\rho}
\ee
The Dirac bracket is defined as
\be
\label{dirac bracket}
\{ A(x), B(x) \}_{DB} = \{ A(x), B(y) \}_{PB} -  \left\{ A(x), \psi_i(z)\right
\} \Delta^{-1}_{i,j}(z,z') \left\{ \psi_j(z'), B(y) \right\}
\ee
where integrations over $z$  and $z'$ are implicit.

We therefore find that the algebra of the remaining currents $J_{2\rho}$ and
$J_{\rho}$ under Dirac bracket yields the $N=1$ super conformal algebra if we
define \br
L(x) = {\lambda \over kB_{2\rho}} J_{2\rho} \;;\;\;\;\;
G(x) = {\left ({{2\lambda \over kB_{2\rho}
\varepsilon (\rho, \rho) }}\right )}^{1\over 2}J_{\rho}.
\er
In other words, the Dirac bracket yields
\br
\label{N=1 a}
{[L(x), L(y)]}_{DB} = 2 L(y) \delta'(x-y) - L'(x) \delta (x-y) - {k\over 4}
\delta'''(x-y)   \nonumber
\er
\br
\label{N=1 b}
{[L(x), G(y)]}_{DB} = {3\over 2} G(y) \delta'(x-y) - G'(x) \delta (x-y)
\er
\br
\label{N=1 c}
{[G(x), G(y)]}_{DB} = 2 L(y) \delta (x-y) - k \delta''(x-y) \nonumber
\er
where we have used the following relations
\br
\varepsilon (2\rho, -\rho) B_{\rho} =
\varepsilon (\rho, \rho) B_{2\rho}
\er
and
\br
2 B_{\rho} = - {\varepsilon (2\rho, -\rho)}^2. \nonumber
\er
which were obtained from the Super Jacobi identities and the symmetry
properties of the brackets.

Moreover, solving the
constraints and gauge fixings, $\psi_i(x) = 0$ , $i = 1, 2, 3 $ for $L(x)$  and
$G(x)$, we find an explicit realization of the algebraic structure (\ref{N=1
a}) in terms of the Super Toda (Super Liouville) fields, i.e.,
\br
L(x) =  \left \{ {1\over 2}(\pa_+{\phi})^2
- {1\over 2} \pa_+^2{\phi} - {1\over 2}\lambda \varepsilon (\rho, \rho)\xi
\pa_+ {\xi} \right \}
\er
\br
G(x) = k B_{\rho}\sqrt {{2\lambda \over k B_{2\rho} \varepsilon (\rho, \rho)}}
\left\{\xi \pa_+{\phi} - \pa_+ {\xi} \right\}.
\er
Such realization can be done if the Dirac bracket is replaced by the
canonical  Poisson bracket derived from the super Toda action,
\cite{Hollowood}.
\subsection{{\cal G} = OSP(2,2) and the N = 2 Super Conformal Theory}

In this case, the even and odd roots are respectively $\pm 2\rho$ and
$\pm \rho \pm \epsilon$, $\;\;({\rho}^2=1, \;\;
{\epsilon}^2 = -1)$. The Lie algebra is
\br
[H_{\rho}, E_{\pm \a} ] = {\pm}\rho \cdot \a E_{\pm \a}  \; ;
 \;\; \;
[H_{\epsilon}, E_{\pm \a} ] = {\pm}\epsilon \cdot \a E_{\pm \a}
 \; ;\;\;for\;\;\;\a = 2\rho,\;\; (\rho + \epsilon), \;\;(\rho - \epsilon)\;;
\er
\br
[H_{\rho}, H_{\epsilon}] = 0\;;\;\;\;
[E_{+ 2\rho}, E_{-2\rho}] = 2 B_{2\rho} H_{\rho} \;;\;\;\;
\{ E_{+\rho + \epsilon}, E_{-\rho - \epsilon}\}= B_{\rho+\epsilon}
(H_{\rho} + H_{\epsilon}) \nonumber
\er
\br
\{ E_{+\rho -\epsilon}, E_{-\rho +\epsilon}\}= B_{\rho-\epsilon}
(H_{\rho} - H_{\epsilon})\;;\;\;\;
[E_{\pm 2\rho}, E_{\mp {(\rho + \epsilon)}} ] =
\varepsilon(\pm 2\rho,\mp(\rho + \epsilon)) E_{\pm {(\rho -\epsilon)}}
\nonumber
\er
\br
[E_{\pm 2\rho}, E_{\mp {(\rho - \epsilon)}} ] =
\varepsilon(\pm 2\rho,\mp(\rho - \epsilon)) E_{\pm {(\rho +\epsilon)}}\; ; \;\;
   \;
\{ E_{\pm {(\rho + \epsilon)}}, E_{\pm {(\rho - \epsilon)}}\} =
\varepsilon(\pm (\rho+ \epsilon), \pm (\rho - \epsilon)) E_{\pm 2\rho}
\nonumber
\er
\br
[E_{\pm 2\rho}, E_{\pm {(\rho \pm \epsilon)}}]
=\{ E_{\pm {(\rho + \epsilon)}}, E_{\mp {(\rho - \epsilon)}}\} =   \{ E_{\pm
{(\rho + \epsilon)}}, E_{\pm {(\rho + \epsilon)}}\} =  \{ E_{\pm {(\rho -
\epsilon)}}, E_{\pm {(\rho - \epsilon)}}\} = 0 \nonumber
\er

Following Eq. (\ref{gauss components}) we define the bosonic and fermionic
fields as
\be \cf_{+} = \zeta_1 (x) E_{\rho + \epsilon} + \zeta_2 (x) E_{\rho -
\epsilon} + f(x) E_{2\rho} \;;\;\;\;   \cf_{-} = \xi_1 (x) E_{-\rho -
\epsilon} + \xi_2(x) E_{-\rho + \epsilon} + g(x) E_{-2\rho}
\ee
and two extra bosonic fields, $\phi_1(x)$ and $\phi_2(x)$,  associated
to the Cartan subalgebra.  The  components
of the current are the following, (with the normalization $Str H_{\rho}^2 =
- Str H_{\epsilon}^2 = 1$),
\br
{1\over k}J_{-2\rho} = B_{2\rho}\left \{\pa_+f -
{\varepsilon(\rho+\epsilon,\rho-\epsilon)\over 2} (\zeta_1 \pa_+\zeta_2  +
\zeta_2 \pa_+ \zeta_1)\right \} e^{-2\phi_1}
\er
\br
{1\over k}J_{H_{\rho}} = \pa_+ \phi_1 + {2  \over k}g J_{-2\rho}
-  B_{\rho+\epsilon}\xi_1 \pa_+ \zeta_1 e^{-\phi_1 + \phi_2}
-  B_{\rho-\epsilon} \xi_2 \pa_+ \zeta_2 e^{-\phi_1-\phi_2} \nonumber
\er
\br
{1\over k}J_{H_{\epsilon}} = -\pa_+ \phi_2 +
{\varepsilon(\rho+\epsilon,\rho-\epsilon) \over k}
\xi_1 \xi_2 J_{-2\rho} + B_{\rho +\epsilon} \xi_1 \pa_+ \zeta_1 e^{-\phi_1
+\phi_2} - B_{\rho -\epsilon} \xi_2 \pa_+ \zeta_2 e^{-\phi_1-\phi_2} \nonumber
\er
\br
{1\over k}J_{2\rho}  = & B_ {2\rho} & \left \{-{2\over k}g^2 J_{-2\rho}
+ \pa_+ g
-{\varepsilon(\rho+\epsilon,\rho-\epsilon) \over 2} (\xi_1 \pa_+ \xi_2 +
 \xi_2 \pa_+ \xi_1) -2 g \pa_+ \phi_1  + \right. \nonumber \\
&  & \mbox{}+ \left. B_{\rho + \epsilon}g \xi_1 \pa_+ \zeta_1
e^{-\phi_1 + \phi_2}  +  B_{\rho - \epsilon} g \xi_2 \pa_+ \zeta_2 e^{-\phi_1 -
\phi_2} + \varepsilon(\rho+\epsilon,\rho-\epsilon) \xi_1 \xi_2 \pa_+ \phi_2
\right \}  \nonumber
\er
\br
{1\over k}J_{\rho+\epsilon}& = & B_{\rho+\epsilon} \left \{{2  \over k}
g \xi_1 J_{-2\rho} -
(\varepsilon (2\rho ,-\rho +\epsilon )g + B_{\rho - \epsilon}\xi_1 \xi_2) \pa_+
\zeta_2  e^{-\phi_1 - \phi_2} - \pa_+ \xi_1\right. \\ \nonumber
& & \mbox{} \left. +  \xi_1 (\pa_+ \phi_1 - \pa_+ \phi_2)\right \} \nonumber
\er
\br
{1\over k}J_{\rho -\epsilon} & = & B_{\rho-\epsilon} \left \{
{2  \over k} g \xi_2 J_{-2\rho} -
(\varepsilon(2\rho,-\rho+\epsilon)  g + B_{\rho+\epsilon}
\xi_2 \xi_1) \pa_+ \zeta_1 e^{-\phi_1 + \phi_2} - \pa_+ \xi_2 \right. \\
\nonumber
&  & \mbox{} \left.+\xi_2 (\pa_+ \phi_1 + \pa_+ \phi_2) \right \} \nonumber
\er
\br
{1\over k}J_{-(\rho+\epsilon)} = B_{\rho+\epsilon}\left \{
{\varepsilon (2\rho,-\rho +\epsilon ) \over k B_{2\rho}} \xi_2 J_{-2\rho} +
\pa_
   +
\zeta_1  e^{-\phi_1 + \phi_2}\right \} \nonumber
\er
\br
{1\over k}J_{-(\rho - \epsilon)} = B_{\rho-\epsilon} \left \{
{\varepsilon(2\rho,-\rho-\epsilon) \over k B_{2\rho} }\xi_1 J_{-2\rho} +
\pa_+ \zeta_2 e^{-\phi_1 - \phi_2} \right \}  \nonumber
\er
The constraints together with their respective gauge fixings are the following,
\br
\psi_1 = J_{-2\rho} - \lambda \;;\;\; \psi_2 = J_{H_\rho} \;;\;\;
\psi_3 = J_{-(\rho + \epsilon)} \;;\;\;\psi_4 = J_{-(\rho - \epsilon)}.
\er

We now take the remaining currents $J_{2\rho}$, $J_{H_{\epsilon}}$,
$J_{\rho + \epsilon}$ and $J_{\rho - \epsilon}$ and make the following
combination
\br
L(x) = {\lambda \over kB_{2\rho}} J_{2\rho} - {1\over 2k} J_{H_{\epsilon}}^2
\;;\;\;\;\;\;\; T(x) = J_{H_\epsilon}
\er
\br
G_+(x) =\sqrt {{2\lambda \over k B_{2\rho} \varepsilon(\rho +\epsilon, \rho
-\epsilon)  }} J_{\rho - \epsilon}
\;;\;\;\;\;\;\;
G_-(x) =\sqrt {{2\lambda \over kB_{2\rho}
\varepsilon(\rho +\epsilon, \rho-\epsilon) }} J_{\rho + \epsilon}
\er
The Dirac brackets evaluated with these operators give us the interesting
case of N=2 super conformal algebra, (see \cite{Schwimmer})  i.e.,
\br
\label{N=2 algebra}
{[L(x), L(y)]}_{DB} = 2 L(y) \delta'(x-y) - L'(x) \delta (x-y) - {k\over 4}
\delta'''(x-y)   \nonumber
\er
\br
{[L(x), G_{\pm}(y)]}_{DB} = {3\over 2} G_{\pm}(y) \delta'(x-y) -
G_{\pm}'(x) \delta (x-y)
\er
\br
{[L(x), T(y)]}_{DB} = T(y) \delta'(x-y) - T'(x) \delta (x-y)
\er
\br
{[T(x), T(y)]}_{DB} = - k \delta'(x-y)
\er
\br
{[T(x), G_{+}(y)]}_{DB} = G_+(y)\delta (x-y)
\er
\br
{[T(x), G_{-}(y)]}_{DB} = - G_-(y) \delta (x-y)
\er
\br
{[G_-(x), G_+(y)]}_{DB} = 2 L(y) \delta (x-y) + 2T(y)\delta'(x-y)
- T'(x) \delta (x-y) - k \delta''(x-y)
\er
\br
{[G_+(x), G_+(y)]}_{DB} = [G_-(x), G_-(y)]_{DB} = 0
\er
where we have used the following relations
\br
\varepsilon (2\rho, -\rho +\epsilon) B_{\rho + \epsilon} =
\varepsilon (\rho +\epsilon, \rho -\epsilon) B_{2\rho} =
\varepsilon (2\rho, -\rho -\epsilon) B_{\rho - \epsilon}
\er
and
\br
\varepsilon (2\rho, -\rho -\epsilon)\varepsilon (2\rho, -\rho +\epsilon) =
-2 B_{2\rho} \nonumber
\er
which were obtained from the Super Jacobi identities.

Again, solving the constraints and gauge fixings for the remaining currents
we finally obtain the realization of the generators of the $N=2$ super
conformal algebra in terms of the Toda fields, given by
\br
L(x) = \left\{{1\over 2}(\pa_+{\phi_1})^2 - {1\over 2}\pa_+^2{\phi_1} -
{1\over 2}(\pa_+{\phi_2})^2 - {1\over 2}\varepsilon(\rho +\epsilon,
\rho-\epsilon)
\lambda (\xi_1 \pa_+ {\xi_2} + \xi_2 \pa_+ {\xi_1}) \right\}  \nonumber
\er
\br
T(x) = k\left \{- \pa_+ \phi_2 - \varepsilon(\rho +\epsilon, \rho-\epsilon)
\lambda \xi_1 \xi_2\right \}
\er
\br
G_-(x) = k \sqrt {{2 \lambda \over k\varepsilon(\rho +\epsilon, \rho-\epsilon)
B_{2\rho}}} B_{\rho + \epsilon}
\left\{ - \pa_ +\xi_1 +  \xi_1(\pa_+{\phi_2} - \pa {\phi_1}) \right\}
\nonumber \er
\br
G_+(x) = k \sqrt {{2\lambda \over k \varepsilon(\rho +\epsilon,
\rho-\epsilon)}}
B_{\rho + \epsilon}
\left\{ - \pa_+ \xi_2 + \xi_2(\pa_+{\phi_1} + \pa_+{\phi_2}) \right\}
\nonumber
\er
where the bracket in equations (\ref{N=2 algebra}) is replaced by the
Poisson brackets, derived from the action of the model, \cite{Hollowood}.
\subsection{{\cal G} = OSP(3,2) and the Super W-Algebra}

The novelty of this example is the existence of a operator of conformal
spin $5/2$ leading to a super W-algebra as an extension of the $N=1$ super
conformal structure already discussed in detail in sections $5.1$. As
described in section $4$, the even and the odd roots are
respectively,
\br
\Delta _0 = \lbrace \pm \epsilon , \pm 2\rho \rbrace \nonumber
\er
\be
\Delta _1 = \lbrace \pm \rho, \pm \rho \pm \epsilon \rbrace.
\nonumber
\ee
The constraints and gauge fixings for this case are chosen to be,
\br
\psi_1 = J_{-2\rho} - \lambda  \;;\;\;\; \psi_2 = J_{H_{\rho}} \;;\;\;\;
\psi_3= J_{-\epsilon} - \mu\;;\;\;;\psi_4 = J_{H_{\epsilon}} \;;\;\;\;\nonumber
\\
 \psi_5 = J_{-\rho}\;;\;\;\;\psi_ 6 = J_{\rho - \epsilon}\;;\;\;\;
\psi_7 = J_{-\rho - \epsilon}\;;\;\;\; \psi_8 = J_{-\rho +\epsilon}\;\;\;\;
\er
It was also argued that there are two bosonic remaining currents of
conformal spin $2$ namely $J_{\epsilon}(x)$ and $J_{2\rho}(x)$ and two
fermionic of conformal spin $3/2$ and $5/2$ corresponding to $J_{\rho}(x)$
and $J_{\rho+\epsilon}(x)$, respectively.

We now take the remaining currents $J_{2\rho}$, $J_{\epsilon}$, $J_{\rho}$,
$J_{\rho +\epsilon}$ in the following combinations
\br
\label{algebra 3,2}
L(x) = {\lambda \over kB_{2\rho} } J_{2\rho} +
{\mu \over k B_{\rho}} J_{\epsilon}
\er
\br
R(x) = {\lambda \over k B_{2\rho}} J_{2\rho}
- {\mu \over kB_{\rho}} J_{\epsilon}
\er
\br
G(x) =\sqrt{\left ({{2\lambda \over kB_{2\rho}
\varepsilon (\rho, \rho) }}\right )}\;\; J_{\rho}
\er
\br
W_{5\over 2} = {\mu \over 2 k \varepsilon (\rho+\epsilon,
-\epsilon)}\sqrt{{2\lambda \over k  B_{2\rho}\varepsilon(\rho,\rho)}}\;\;
J_{\rho+\epsilon}
\er
and evaluate the Dirac brackets. The resulting algebra is given by
\br
\label{N=1 osp32 algebra}
{[L(x), L(y)]}_{DB} = 2 L(y) \delta'(x-y) - L'(x) \delta (x-y) + {3k\over 4}
\delta'''(x-y)
\er
\br
\label{N=1 osp32b algebra}
{[L(x), G(y)]}_{DB} = {3\over 2} G(y) \delta'(x-y) - G'(x) \delta (x-y)
\er
\br
\label{N=1 osp32c algebra}
{\{G(x), G(y)\}}_{DB} = 2 L(y) \delta(x-y) + 3 k \delta''(x-y)
\er
Equations (\ref{N=1 osp32 algebra})-(\ref{N=1 osp32c algebra}) provide the
realization of $N=1$ super conformal algebra. The second spin $2$ and the
spin $5/2$ currents transform, under the Virasoro generator, as
\br
{[L(x), R(y)]}_{DB} = 2 R(y) \delta'(x-y) - R'(x) \delta (x-y) - {5k\over 4}
\delta'''(x-y)
\er
\br
{[L(x), W_{5\over 2}(y)]}_{DB} = {5\over 2} W_{5\over 2}(y)\delta'(x-y) -
W'_{5\over 2}(y)\delta (x-y) - {1\over 2} G(y) \delta''(x - y)
\er
Notice that the spin $5/2$ current $W_{5\over 2}(x)$ does not transform
as a primary field. However, the linear combination $ W_{5\over 2}(x) -
{1\over 3} G'(x)$ does. The remaining Dirac brackets we obtain are
\br
{[R(x), R(y)]}_{DB}& = &2 L(y) \delta'(x-y) - L'(x) \delta (x-y) + {3k\over 4}
\delta'''(x-y)  \\
{[R(x), G(y)]}_{DB}& = &{3\over 2} G(y) \delta'(x-y) - G'(x) \delta (x-y)
 -4 W_{5\over 2}(x) \delta(x-y) \\
{[R(x), W_{5\over 2}(y)]}_{DB} & = & {1\over 2} W_{5\over 2}(y)\delta'(x-y) -
W'_{5\over 2}(y)\delta (x-y)  \nonumber \\
 & & \mbox{} - {\mu \over k} G(y)\left ( L(y) + R(y)\right )
\delta (x-y)  \nonumber \\
&  & \mbox{} +\left (G(x) + {1\over 2} G(y) \right )\delta''(x-y) \\
{\{G(x), W_{5\over 2}(y)\}}_{DB} & = & \left (L(y) + R(y) \right )\delta'(x-y)
+{1\over 4} \left ( L'(x) - R'(x) \right ) \delta(x-y)  \nonumber \\
&  & \mbox{} - k\delta'''(x-y) \\
{ \{ W_{5\over 2}(x), W_{5\over 2}(y)\} }_{DB} & = &
- {1\over 2k} \left ( L^2(x) + L(x) R(x)\right ) \delta (x-y)
 + {3\over 8k}\left ( L(y) + L(x) \right )  \delta''(x-y) \nonumber \\
&  & \mbox{} + {1\over 8k}\left ( R(y) + R(x) \right )  \delta''(x-y) +
{1\over 4k} G'(x) G(x)\delta (x-y)  \nonumber \\
&  & \mbox{} + {1\over k} W_{5\over 2}(x)G(x) \delta(x-y) -{k\over 4}
\delta''''(x-y).
\er
The relations from the Super Jacobi identities we have used are
\br
\varepsilon (2\rho, -\rho +\epsilon) B_{\rho + \epsilon} =
\varepsilon (\rho +\epsilon, \rho -\epsilon) B_{2\rho} =
\varepsilon (2\rho, -\rho -\epsilon) B_{\rho - \epsilon} \nonumber
\er
\br
\varepsilon (\rho +\epsilon, -\rho) B_{\epsilon} =
- \varepsilon (\rho +\epsilon, -\epsilon) B_{\rho} =
\varepsilon( \epsilon, \rho) B_{\rho + \epsilon} \nonumber
\er
\br
\varepsilon (\rho -\epsilon, -\rho) B_{\epsilon} =
\varepsilon (-\rho, \epsilon) B_{\rho - \epsilon} =
- \varepsilon (\rho -\epsilon, \epsilon) B_{\rho}
\er
\br
\varepsilon (2\rho,-\rho -\epsilon)\varepsilon (2\rho,-\rho +\epsilon) =
- {\varepsilon (2\rho, -\rho)}^2 = -2B_{2\rho}  \nonumber
\er
\br
\varepsilon(\rho, \rho) B_{2\rho} =\varepsilon(2\rho, -\rho) B_{\rho}
\;;\;\;\;
\varepsilon(\rho-\epsilon, -\rho) \varepsilon(\epsilon, -\rho) = B_{\rho}.
\nonumber
\er
At this point we can use the Gauss decomposition to obtain the components of
the
current.  Again, solving the appropriate constraints and gauge fixings for the
remaining currents we find, in particular, another representation for the $N=1$
super conformal algebra.

\subsection{\lie = $D(2,1 \vert \a)$ and the Super W-Algebra}

This example presents three non commuting Virasoro generators, two fields of
conformal spin $3/2$ and one of spin $5/2$. We show that they lead to two non
commuting $N=1$ super W-algebras.

The even and the odd roots are respectively,
\br
\Delta _0 & = & \lbrace \pm 2\epsilon_i \rbrace \;, i = 1, 2, 3 \nonumber \\
\Delta _1 & = & \lbrace \pm \epsilon_1 \pm \epsilon_2 \pm \epsilon_3 \rbrace
\er
The dimension of this group is 17. The simple (fermionic) roots are
\br
\a_1 = - \epsilon_1 + \epsilon_2 + \epsilon_3\;;\;\;
\a_2 =  \epsilon_1 - \epsilon_2 + \epsilon_3 \;;\;\;
\a_3 =  \epsilon_1 + \epsilon_2 - \epsilon_3
\er
The other positive roots are
\br
\begin{array}{ll}
\a_4  =   \a_1 + \a_2 = 2\epsilon_3\;;\;\;\;
&\a_5  =   \a_1 + \a_3 = 2\epsilon_2\;;\;\;\; \\
\a_6  =  \a_2 + \a_3 = 2\epsilon_1\;;\;\;\;
&\a_7  =   \a_1 + \a_2 + \a_3 = \epsilon_1 + \epsilon_2 + \epsilon_3
\end{array}
\er

The constraints and corresponding gauge fixings are
\br
\begin{array}{lll}
\psi_1 = J_{-\a_1}\; ; \; \; \; &\psi_2 =J_{-\a_2}\; ; \; \; \;
&\psi_3 = J_{-\a_3} \\
\psi_4 = J_{-\a_4} - \lambda_4\; ; \; \; \; &\psi_5 =  J_{H_3}&  \\
\psi_6 = J_{-\a_5} - \lambda_5 \; ; \; \; \; &\psi_7 = J_{H_2}&  \\
\psi_8 = J_{-\a_6} - \lambda_6\; ; \; \; \; &\psi_9 = J_{H_1} &  \\
\psi_{10} = J_{-\a_7}\; ; \; \; \; &\psi_{11} = J_{\a_3}&
\end{array}
\er

After those are imposed we are left with six remaining currents which will
generate the super W-algebra. We have chosen the following normalization for
them
\br
\label{3l}
L_1(x) = {1\over k} {\lambda_4 \over B_{\a_4}}J_{\a_4}(x)\;;\;\;\;
L_2(x) = {1\over k} {\lambda_5 \over B_{\a_5}}J_{\a_5}(x)\;;\;\;\;
L_3(x) = {1\over k} {\lambda_6 \over B_{\a_6}}J_{\a_6}(x)\;\;\;\;
\er
\br
\label{3/21}
W^{(1)}_{3\over 2}(x) & = & \sqrt{(\a_1 \cdot \a_3)
{\lambda_4 \lambda_5 \over k\lambda_6}
{\varepsilon(\a_1,\a_2) \over \varepsilon(\a_1,\a_3)\varepsilon(\a_2,\a_3)}
{B_{\a_3} \over B_{\a_1}B_{\a_5} }}\;\;
 J_{\a_1}(x)  \\
\label{3/22}
W^{(2)}_{3\over 2}(x) & = & \sqrt{(\a_{2} \cdot \a_{3})
{\lambda_4 \lambda_6 \over k\lambda_5}
{\varepsilon(\a_1,\a_2) \over \varepsilon(\a_1,\a_3)
\varepsilon(\a_2,\a_3)}
{B_{\a_3} \over  B_{\a_2}B_{\a_6} }} \;\;
J_{\a_2}(x) \\
\label{5/2}
W_{5\over 2}(x) & = & \sqrt{(\a_1 \cdot \a_3) (\a_2 \cdot \a_3)
{\lambda_4 \lambda_5  \lambda_6 \over k}
{\varepsilon(\a_1,\a_2) \over \varepsilon(\a_1,\a_3)
\varepsilon(\a_2,\a_3)}
{B_{\a_3}\over  B_{\a_5}B_{\a_6}B_{\a_7}} } \;\;
J_{\a_7}(x)
\er
where $B_{\a} \equiv Str(E_{\a} E_{-\a})$.

The structure constants $\varepsilon (\beta , \gamma )$ are determined from the
root system and the super Jacobi identities. However there are some
arbitrariness in the choice of the signs of such  structure constants. We
denote
$\varepsilon (\beta , \gamma ) = \eta (\beta , \gamma ) \mid \varepsilon (\beta
   ,
\gamma ) \mid $, where $\eta (\beta , \gamma ) = \pm 1$. We have chosen
\be
\eta (\a_1 ,\a_2 ) = \eta (\a_1 ,\a_3 ) =  \eta (\a_2 ,\a_3 ) = \eta (\a_1
,\a_6
) = 1
\ee
This choice completely fixes, through the super Jacobi identities, the signs of
all other structure constants $\varepsilon (\beta , \gamma )$.

We parametrize the scalar product on the root space as
\br
\epsilon_1^2 = -{1\over 2} (1 + \a)\; ; \; \;
\epsilon_2^2 = {1\over 2}\; ; \; \;
\epsilon_3^2 = {\a \over 2}\; ; \; \;
\epsilon_i \cdot \epsilon_j = 0\; , \; \; i\not= j
\er
where $\a$ is a parameter and  $\a \neq 0\;,\;-1$. (From now on the $\a 's$
appearing in the formulae are parameters and not roots).

The super W-algebra is given by the Dirac bracket of the remaining currents
(\ref{3l}-\ref{5/2}). It has a very interesting structure as we now discuss.
The currents $L_i (x)$, $i=1,2,3$, given in (\ref{3l}) generate three non
commuting Virasoro algebras
\br
{[L_i(x), L_j(y)]}_{DB} & = & \delta_{i,j} \left \{ 2L_i(y) \delta'(x-y) -
\pa_yL_i(y) \delta(x-y) - c_i \delta'''(x-y)\right \}  \nonumber \\
&   & \mbox{} - {1\over 2k } \left ( \sum_{k=1}^3 \epsilon_{ijk}\right )
W^{(1)}_{3\over 2}(y) W^{(2)}_{3\over 2}(y) \delta(x-y)
\er
where $\epsilon_{ijk}$ is the completely antisymmetric symbol
($\epsilon_{123}=1$), and the central terms are given by
\be
c_1 = {k\over 2} {Str(H_3 H_3) \over \a}\; ; \; \;
c_2 = {k\over 2} Str(H_2 H_2) \; ; \; \;
c_3 = -{k\over 2} {Str(H_1 H_1) \over \( 1 + \a \) }
\ee
Under these Virasoro generators, the spin $3/2$ currents transform as
\br
\label{first N1a}
{[L_2(x), W^{(1)}_{3\over 2}(y)]}_{DB} & = & {3\over 2} W^{(1)}_{3\over 2}(y)
\delta' (x-y) - \pa_y W^{(1)}_{3\over 2}(y) \delta (x-y) \\
\label{decoupling a}
{[L_1(x) + L_3(x), W^{(1)}_{3\over 2}(y)]}_{DB} & = & 0 \\
{[L_1(x) - L_3(x),W^{(1)}_{3\over 2}(y)]}_{DB} & = & W^{(1)}_{3\over 2}(y)
\delta' (x-y) + {2\over k^2} W_{5\over 2}(y) \delta (x-y) \nonumber \\
&   & \mbox{} - 2 \left( W^{(2)}_{3\over 2}(y) \delta' (x-y) - \pa_y
W^{(2)}_{3\over 2}(y) \delta (x-y) \right)
\er
and
\br
\label{second N1a}
{[L_3(x), W^{(2)}_{3\over 2}(y)]}_{DB} & = & {3\over 2} W^{(2)}_{3\over 2}(y)
\delta' (x-y) - \pa_y W^{(2)}_{3\over 2}(y) \delta (x-y) \\
\label{decoupling b}
{[L_1(x) + L_2(x), W^{(2)}_{3\over 2}(y)]}_{DB} & = & 0 \\
{[L_1(x) - L_2(x), W^{(2)}_{3\over 2}(y)]}_{DB} & = & W^{(2)}_{3\over 2}(y)
\delta' (x-y) + {2\over k^2} W_{5\over 2}(y) \delta (x-y) \nonumber \\
&   & \mbox{} - 2 \left( W^{(1)}_{3\over 2}(y)
\delta' (x-y) - \pa_y W^{(1)}_{3\over 2}(y) \delta (x-y) \right)
\er
The spin $5/2$ current transforms as
\br
{[L_1(x), W_{5\over 2}(y)]}_{DB} & = & {3\over 2} W_{5\over 2}(y)
\delta'(x-y) - \pa_y W_{5\over 2}(y) \delta(x-y)   \nonumber \\
&   &  \mbox{} + {1\over 2} \left \{ \a L_1(y) - L_2(y) - (1 + \a) L_3(y)
\right \} W^{(1)}_{3\over 2}(y) \delta (x-y) \nonumber \\
&   &  \mbox{} + {1\over 2}\left \{ \a L_1(y) + L_2(y) + (1 + \a) L_3(y)
\right \} W^{(2)}_{3\over 2}(y) \delta (x-y) \nonumber \\
&   &  \mbox{}- {k\over 2} \left \{
\left ( W^{(1)}_{3\over 2}(y) + W^{(2)}_{3\over 2}(y)\right )\delta'' (x-y)
 \right. \nonumber \\
&   &  \mbox{} - 2 \pa_y \left( W^{(1)}_{3\over 2}(y) + W^{(2)}_{3\over2}(y)
\right)  \delta' (x-y)   \nonumber \\
&  & \mbox{}  \left. + \pa^2_y \left(
W^{(1)}_{3\over 2}(y) + W^{(2)}_{3\over 2}(y) \right) \delta (x-y) \right
\} \\
{[L_2(x),  W_{5\over 2}(y)]}_{DB} & = &{1\over 2} W_{5\over 2}(y) \delta' (x-y)
   -
L_2(y) W^{(2)}_{3\over 2}(y) \delta (x-y) \nonumber \\
&   & \mbox{} - {1\over 2}\left \{ \a L_1(y) - L_2(y) + (1 + \a) L_3(y)
\right \} W^{(1)}_{3\over 2}(y) \delta (x-y) \nonumber \\
&   & \mbox{} + {k\over 2} \left \{ W^{(1)}_{3\over 2}(y)\delta'' (x-y) -2
\pa_y W^{(1)}_{3\over 2}(y)  \delta' (x-y) + \pa^2_y W^{(1)}_{3\over 2}(y)
\delta (x-y) \right \} \nonumber \\
&   & \mbox{} + c_2 W^{(2)}_{3\over 2}(y)\delta'' (x-y)
\\
{[L_3(x),  W_{5\over 2}(y)]}_{DB} & = &{1\over 2} W_{5\over 2}(y) \delta' (x-y)
   +
(1 + \a ) L_3(y)W^{(1)}_{3\over 2}(y) \delta (x-y) \nonumber \\
&   & \mbox{} -{1\over 2} \left \{ \a L_1(y) - L_2(y) + (1 + \a) L_3(y)
\right \} W^{(2)}_{3\over 2}(y) \delta (x-y) \nonumber \\
&   & \mbox{} +{k\over 2}\left \{ W^{(2)}_{3\over 2}(y)\delta'' (x-y) -2
\pa_y W^{(2)}_{3\over 2}(y)  \delta' (x-y) + \pa^2_y W^{(2)}_{3\over 2}(y)
\delta (x-y) \right \} \nonumber \\
&   & \mbox{} - c_3\( 1 + \a \) W^{(1)}_{3\over 2}(y)\delta'' (x-y)
\er

Notice that $W^{(1)}_{3\over 2}(x)$  and $W^{(2)}_{3\over 2}(x)$ are  primary
field (of spin $3/2$) with respect to the diagonal Virasoro generator $L_1(x)
+ L_2(x) + L_3(x)$. $W_{5\over 2}(x)$, on the other hand, is not a primary
field. This is a consequence of the gauge we are using. However one can check
that
\be
V_{5\over 2}(x) \equiv W_{5\over 2}(x) + {k\over 3} \pa_x \( STr(H_1 H_1 )
W^{(1)}_{3\over 2}(x) + STr(H_2 H_2 ) W^{(2)}_{3\over 2}(x) \right)
\ee
is a primary field of spin $5/2$ w.r.t. the diagonal Virasoro $L_1(x) + L_2(x)
+
L_3(x)$.

The Dirac brackets involving spin $3/2$ currents are
\br
\label{first N1b}
{\{ W^{(1)}_{3\over 2}(x), W^{(1)}_{3\over 2}(y)\}}_{DB} & = & -2 L_2(y)
\delta (x-y) + 2k \delta'' (x-y)  \\
\label{second N1b}
{\{ W^{(2)}_{3\over 2}(x), W^{(2)}_{3\over 2}(y)\}}_{DB} & = & 2 (1+\a ) L_3(y)
   \delta
(x-y) + 2k \delta'' (x-y)  \\
{\{ W^{(1)}_{3\over 2}(x), W^{(2)}_{3\over 2}(y)\}}_{DB} & = & \left \{-\a
L_1(y
   ) +
L_2(y) - (1 + \a ) L_3(y) \right \} \nonumber \\
&   & \mbox{} - \delta (x-y) - k \delta'' (x-y)
\er
We also have
\br
{\{ W^{(1)}_{3\over 2}(x), W_{5\over 2}(y)\}}_{DB} & = & -{1\over 2}
W^{(1)}_{3\over 2}(y)  W^{(2)}_{3\over 2}(y) \delta (x-y) +
 k^2 \delta''' (x-y) \nonumber \\
&   & \mbox{} + k\left \{ (1+\a)\left (\a L_1(y) -(1+\a) L_3(y)
\right)\delta' (x-y)  \right. \nonumber \\
&   & \mbox{}  \left. - \left ( L_2(y)\delta' (x-y) - \pa_y L_2(y)
\delta (x-y) \right ) \right \}
\\
{\{ W^{(2)}_{3\over 2}(x), W_{5\over 2}(y)\}}_{DB} & = & {1\over 2}(1+\a)
W^{(1)}_{3\over 2}(y)  W^{(2)}_{3\over 2}(y) \delta (x-y) + k^2
\delta''' (x-y)  \nonumber \\
&   & \mbox{} + k \left \{ \left ( -\a L_1(y) +  L_2(y) \right ) \delta' (x-y)
\right. \nonumber\\
&   & \mbox{}  \left. +   (1+\a) \left ( L_3(y) \delta'(x-y) -\pa_y L_3(y)
\delta (x-y)\right ) \right \}
\er
and
\br
{\{ W_{5\over 2}(x), W_{5\over 2}(y)\}}_{DB} & = & \left \{ -(1+\a )
W^{(1)}_{3\over 2}(y) +
 W^{(2)}_{3\over 2}(y) \right \} W_{5\over2}(y) \delta (x-y) \nonumber \\
&   & \mbox{} - {k\over 2} {\left \{ \a L_1(y) - L_2(y) + (1+\a ) L_3(y)
\right \}}^2 \delta (x-y)  \nonumber \\
&   & \mbox{} -2k (1+\a ) L_2(y)L_3(y)\delta (x-y)  \nonumber \\
&   & \mbox{} + {k^2\over 2} \left \{ \a ( L_1(x) +
L_1(y)) - ( L_2(x) + L_2(y))   \right. \nonumber \\
&   & \mbox{} \left.+ (1+\a ) (L_3(x) + L_3(y)) \right \}\delta'' (x-y)
- {k^3\over2} \delta'''' (x-y)
\er

Notice, from (\ref{first N1a}) that $W^{(1)}_{3\over 2}(x)$ is a
spin $3/2$ primary field w.r.t $L_2(x)$. From (\ref{first N1b}) we see that the
operators $L_2(x)$ and $W^{(1)}_{3\over 2}(x)$ generate a $N=1$
superconformal subalgebra of the above super $W$-algebra. Analogously, from
(\ref{second N1a}) and (\ref{second N1b}) we see that $L_3(x)$ and
$W^{(2)}_{3\over 2}(x)$ generate another $N=1$ superconformal subalgebra.
However these two subalgebras do not commute.

\section{Conclusions}

In this paper we have achieved a systematic method of classifying and dealing
with the symmetries of the supersymmetric Toda system.  This was accomplished
by exploiting the Hamiltonian reduction of the WZNW model associated to a
super Lie algebra whose simple roots are all fermionic.  This process leads
to an improved energy-momentum tensor of the theory giving rise to
higher spin generators.  Their classification generalizes
the one employed for the bosonic case (\cite {balog}).  Each generator
corresponds to a highest weight state of a representation of a special
$Sl(2)$ subalgebra of $\lie $ selected by the constraints.

The second main point we should highlight relies upon the Gauss decomposition
formula which allows  explicit constructions of the current components in terms
of the WZNW fields.  After imposing the constraints and gauge fixings,  all
degrees of freedom beyond the Super Toda fields are eliminated and the
remaining currents, in particular, provides, in a systematic manner,
representations of $N=1$ and $N=2$ super conformal algebras.

An interesting point we intend to pursue further concerns the construction of
representations of super conformal algebras with higher supersymmetries whose
algebraic structure was discussed in \cite{Schwimmer}.  This involves
fermionic central extensions and may be related to other conformally invariant
models beyond the Super Toda.

\end{document}